\documentclass{article}
\begin{document}

\centerline{{\LARGE Impossible to conjugate or reflect }} %
\centerline{{\LARGE an unknown quantum state via a unitary evolution }}

\centerline{Dafa Li}
\centerline{Department of Mathematical Sciences, Tsinghua
University, Beijing 100084, CHINA} \centerline{email: lidafa@tsinghua.edu.cn}
\centerline{Corresponding author: Dafa Li}

Abstract.

The no-cloning theorem, the no-deleting theorem, and other no-go theorems
play a crucial role in quantum mechanics and quantum information. In this
paper, we show that it is impossible to conjugate or reflect an unknown
quantum state or conjugate or reflect the bit-flip (the phase-flip) of an
unknown state via a universally unitary evolution. We also indicate that the
partial conjugating operations are not anti-unitary and it is impossible to
universally partially conjugate an unknown state via a unitary evolution. We
compute the fidelities of the optimal physical approximations to conjugating
and conjugating the bit-flip and phase-flip operations.


\section{Introduction}

W. K. Wootters and W. H. Zurek in 1982 proposed the quantum no-cloning
theorem \cite{Wootters}. The theorem states that it is impossible to create
an identical copy of an unknown quantum state. The theorem plays a crucial
role in the development of quantum computing, quantum teleportation, quantum
cryptography, and quantum key distribution.\

Following no-cloning theorem, Yao gave a physically natural definition of
cloning in the context of classical mechanics using symplectic geometry \cite%
{Yao}.\ Yamaguchi and Kempf showed that encrypted cloning of unknown quantum
states is possible \cite{Yama}

Pati and Braunstein's no-deleting theorem states that the deleting machine
cannot delete one of two identical copies of an unknown quantum state \cite%
{Pati}. No-deleting principle provides intrinsic security to quantum files
in a quantum computer. The relation between no-cloning and no-deleting
principles is investigated in \cite{Qiu, sk, IC, Chen}.

No-hiding theorem and no-broadcasting were proposed in \cite{Samuel} and
\cite{Barnum}, respectively. No-hiding theorem sheds new light on the
black-hole information paradox \cite{Samuel}. Girling et al. gave a simple
formulation of no-cloning and no-hiding \cite{Matt}.

No-flip theorem states that no universal unitary operator can take an
arbitrary unknown qubit to its orthogonal state \cite{bu, bu00}. The
relation between no-flipping theorem and no-partial-erasure theorem is
explored in \cite{Pati05}.

No partial erasure of quantum information and impossibility of partial
swapping of quantum information were studied in \cite{Pati05, IC-2},
respectively.

In this paper, we show that it is impossible to conjugate or reflect an
unknown quantum state or conjugate or reflect the bit-flip (the phase-flip)
of an unknown state via a universally unitary evolution.

In this paper, let $x^{\ast }$ be the complex conjugate of $x$ and let $I$
be the identity. $\sigma _{x}$, $\sigma _{y}$, and $\sigma _{z}$ are
Pauli-operators. $i$ is $\sqrt{-1}$.

\section{No-conjugating}

Suppose there exists a unitary operator $U$ that can operate an arbitrary
unknown state $|\psi \rangle =\alpha |0\rangle +\beta |1\rangle $, where $%
|\alpha |^{2}+|\beta |^{2}=1$, such that

\begin{eqnarray}
U|\psi \rangle &=&|\psi ^{\prime }\rangle ,  \label{in-} \\
|\psi ^{\prime }\rangle &=&p|0\rangle +q|1\rangle  \label{in-0}
\end{eqnarray}%
where $\ $
\begin{equation}
\ p\in \{\alpha ,\alpha ^{\ast }\}\&q\in \{\beta ,\beta ^{\ast }\}\vee p\in
\{\beta ,\beta ^{\ast }\}\&q\in \{\alpha ,\alpha ^{\ast }\}  \label{cd-1}
\end{equation}

By using algebraic method, we directly solve Eqs. (\ref{in-}, \ref{in-0}) to
obtain the unitary operator $U$.

The general form of 2 by 2 unitary operator is
\begin{equation}
U=\left(
\begin{array}{cc}
u & v \\
-e^{i\theta }v^{\ast } & e^{i\theta }u^{\ast }%
\end{array}%
\right) ,  \label{u-mat}
\end{equation}%
where $\theta $ is any real number, $u$ and $v$ are any complex numbers, and
$|u|^{2}+|v|^{2}=1$.

We can rewrite Eq. (\ref{in-}) as
\begin{equation}
U\left(
\begin{array}{c}
\alpha \\
\beta%
\end{array}%
\right) =\left(
\begin{array}{cc}
u & v \\
-e^{i\theta }v^{\ast } & e^{i\theta }u^{\ast }%
\end{array}%
\right) \left(
\begin{array}{c}
\alpha \\
\beta%
\end{array}%
\right) =\left(
\begin{array}{c}
p \\
q%
\end{array}%
\right) .  \label{u-1}
\end{equation}

Expanding Eq. (\ref{u-1}), obtain
\begin{eqnarray}
u\alpha +v\beta &=&p  \label{u-2} \\
e^{i\theta }\beta u^{\ast }-e^{i\theta }\alpha v^{\ast } &=&q  \label{u-3}
\end{eqnarray}

By conjugating, from Eq. (\ref{u-3}) obtain
\begin{equation}
e^{-i\theta }\beta ^{\ast }u-e^{-i\theta }\alpha ^{\ast }v=q^{\ast }
\label{u-4}
\end{equation}

We next solve Eqs. (\ref{u-2}, \ref{u-4}). A calculation yields the
following solution of Eq. (\ref{in-}).
\begin{eqnarray}
u &=&\alpha ^{\ast }p+e^{i\theta }\beta q^{\ast },  \label{u-11} \\
v &=&\beta ^{\ast }p-e^{i\theta }\alpha q^{\ast }.  \label{u-12}
\end{eqnarray}

By \ assigning $p$ and $q$ in Eq. (\ref{cd-1}), we obtain the solutions of
Eq. (\ref{in-}) in Tables 1 and 2 from Eqs. (\ref{u-11}, \ref{u-12}). Here,
a universally unitary operator means that the unitary operator is
independent of an unknown state.

Table 1. The universally unitary operators $U$ for Eq. (\ref{in-})

$%
\begin{tabular}{|l|l|l|l|l|l|l|}
\hline
$p$ & $q$ & $|\psi ^{\prime }\rangle =$ & $\theta $ & $u$ & $v$ & $U$ \\
\hline
$\alpha $ & $\beta $ & $|\Psi _{1}\rangle =\alpha |0\rangle +\beta |1\rangle
$ & $0$ & $1$ & $0$ & $I$ \\ \hline
$\beta $ & $\alpha $ & $|\Psi _{2}\rangle =\beta |0\rangle +\alpha |1\rangle
$ & $\pi $ & $0$ & $1$ & $\sigma _{x}$ \\ \hline
\end{tabular}%
$

Table 2. The non-universally unitary operators $U$ for Eq. (\ref{in-})

$%
\begin{tabular}{|l|l|l|l|l|}
\hline
$p$ & $q$ & $|\psi ^{\prime }\rangle =|\Phi _{i}\rangle $ & $u$ & $v$ \\
\hline
$\alpha $ & $\beta ^{\ast }$ & $|\Phi _{1}\rangle =\alpha |0\rangle +\beta
^{\ast }|1\rangle $ & $e^{i\theta }\beta ^{2}+\alpha \alpha ^{\ast }$ & $%
-e^{i\theta }\alpha \beta +\alpha \beta ^{\ast }$ \\ \hline
$\alpha ^{\ast }$ & $\beta $ & $|\Phi _{2}\rangle =\alpha ^{\ast }|0\rangle
+\beta |1\rangle $ & $e^{i\theta }\beta \beta ^{\ast }+\alpha ^{\ast }\alpha
^{\ast }$ & $-e^{i\theta }\alpha \beta ^{\ast }+\alpha ^{\ast }\beta ^{\ast
} $ \\ \hline
$\alpha ^{\ast }$ & $\beta ^{\ast }$ & $|\Phi _{3}\rangle =\alpha ^{\ast
}|0\rangle +\beta ^{\ast }|1\rangle $ & $e^{i\theta }\beta \beta +\alpha
^{\ast }\alpha ^{\ast }$ & $-e^{i\theta }\alpha \beta +\alpha ^{\ast }\beta
^{\ast }$ \\ \hline
$\beta $ & $\alpha ^{\ast }$ & $|\Phi _{4}\rangle =\beta |0\rangle +\alpha
^{\ast }|1\rangle $ & $e^{i\theta }\alpha \beta +\alpha ^{\ast }\beta $ & $%
-e^{i\theta }\alpha \alpha +\beta \beta ^{\ast }$ \\ \hline
$\beta ^{\ast }$ & $\alpha $ & $|\Phi _{5}\rangle =\beta ^{\ast }|0\rangle
+\alpha |1\rangle $ & $e^{i\theta }\alpha ^{\ast }\beta +\alpha ^{\ast
}\beta ^{\ast }$ & $-e^{i\theta }\alpha \alpha ^{\ast }+\beta ^{\ast }\beta
^{\ast }$ \\ \hline
$\beta ^{\ast }$ & $\alpha ^{\ast }$ & $|\Phi _{6}\rangle =\beta ^{\ast
}|0\rangle +\alpha ^{\ast }|1\rangle $ & $e^{i\theta }\alpha \beta +\alpha
^{\ast }\beta ^{\ast }$ & $-e^{i\theta }\alpha \alpha +\beta ^{\ast }\beta
^{\ast }$ \\ \hline
\end{tabular}%
$

From Table 2, one can see that no matter what is $\theta $, $u$ and $v$ \
are functions of $\alpha $ and $\beta $. That is, no matter what is $\theta $%
, there is no universally unitary operator $U$ that can operate an arbitrary
unknown state $|\psi \rangle =\alpha |0\rangle +\beta |1\rangle $, such that
$U|\psi \rangle =|\Phi _{i}\rangle $, $i=1,...,6$. Thus, obtain six no-go
theorems from $U|\psi \rangle =|\psi ^{\prime }\rangle =|\Phi _{i}\rangle $,
$i=1,...,6$.

From Tables 1 and 2, we have the following theorems.

\textit{Theorem 1}. There exists a universally unitary operator $U$ such
that $U|\psi \rangle =|\psi ^{\prime }\rangle $, where $|\psi \rangle
=\alpha |0\rangle +\beta |1\rangle $, which is an arbitrary unknown state,
and $|\psi ^{\prime }\rangle $ is of the form in Eq. (\ref{cd-1}), if and
only if $|\psi ^{\prime }\rangle =\alpha |0\rangle +\beta |1\rangle $ or $%
\beta |0\rangle +\alpha |1\rangle $.

Clearly, for $|\psi ^{\prime }\rangle =\alpha |0\rangle +\beta |1\rangle $,
the universally unitary operator $U$ is $I$, while for $|\psi ^{\prime
}\rangle =\beta |0\rangle +\alpha |1\rangle $, the universally unitary
operator $U$ is $\sigma _{x}$.

Note that the state $\beta |0\rangle +\alpha |1\rangle $ is called the
bit-flip of the state $\alpha |0\rangle +\beta |1\rangle $. It is easy to
see that $\beta |0\rangle +\alpha |1\rangle =\sigma _{x}(\alpha |0\rangle
+\beta |1\rangle )$.

\textit{Theorem 2 (no-conjugating theorems)}. There is no universally
unitary operator $U$ such that $U|\psi \rangle =|\psi ^{\prime }\rangle $,
where $|\psi \rangle =\alpha |0\rangle +\beta |1\rangle $, which is an
arbitrary unknown state, and $|\psi ^{\prime }\rangle $ is of the form in
Eq. (\ref{cd-1}), if and only if $|\psi ^{\prime }\rangle $ is one of $|\Phi
_{i}\rangle $, $i=1,...,6$ in Table 2. Thus, it yields six no-conjugating
theorems in Table 3.

Table 3. No-conjugating theorems

\begin{tabular}{|l|l|}
\hline
$|\Phi _{1}\rangle $ & no-conjugating the second amplitude \\ \hline
$|\Phi _{2}\rangle $ & no-conjugating the first amplitude \\ \hline
$|\Phi _{3}\rangle $ & no-conjugating \\ \hline
$|\Phi _{4}\rangle $ & no-conjugating the second amplitude of the bit-flip
\\ \hline
$|\Phi _{5}\rangle $ & no-conjugating the first amplitude of the bit-flip \\
\hline
$|\Phi _{6}\rangle $ & no-conjugating the bit-flip \\ \hline
\end{tabular}

For example, let $|\psi ^{\ast }\rangle =\alpha ^{\ast }|0\rangle +\beta
^{\ast }|1\rangle $,\ which is the conjugate state of $|\psi \rangle $. From
Table 2, $|\Phi _{3}\rangle =|\psi ^{\ast }\rangle $. By Theorem 2, we
obtain the no-conjugating theorem, which states that no universally unitary
operator can conjugate an arbitrary unknown quantum state via a unitary
evolution.

Note that the inner product is used for the proof of no-flip theorem while
not for the proof of no-conjugating theorem.

\section{General no-conjugating}

\textit{Lemma 1}. Assume there exists a universally unitary operator $U$
such that $U|\psi \rangle =|\varphi \rangle $, where $|\psi \rangle =\alpha
|0\rangle +\beta |1\rangle $, which is an arbitrary unknown state. Then,
there is a universally unitary operator $W$ such that $W|\psi \rangle
=V|\varphi \rangle $, where $V$ is any universally unitary operator.

From Lemma 1 and Theorem 1, we have the following corollary.

\textit{Corollary 1. }There is a universally unitary operator $U$ such that $%
U|\psi \rangle =V|\psi ^{\prime }\rangle $, where $|\psi \rangle =\alpha
|0\rangle +\beta |1\rangle $, which is an arbitrary unknown state,$\ V$ is
any universally unitary operator, and $|\psi ^{\prime }\rangle $ is one of $%
|\Psi _{1}\rangle $ and $|\Psi _{2}\rangle $ in Table 1.

For example, for an arbitrary unknown state $|\psi \rangle =\alpha |0\rangle
+\beta |1\rangle $, let $V_{1}=diag(-i,i)$, $|\psi ^{\prime }\rangle =$ $%
|\Psi _{2}\rangle $, and $|\phi _{1}\rangle =V_{1}|\psi ^{\prime }\rangle $.
Then, $|\phi _{1}\rangle =-i\beta |0\rangle +i\alpha |1\rangle $. Clearly, $%
\sigma _{y}|\psi \rangle =|\phi _{1}\rangle $.

\textit{Lemma 2.} Assume that there is no universally unitary operator $U$
such that $U|\psi \rangle =|\varphi \rangle $, where $|\psi \rangle =\alpha
|0\rangle +\beta |1\rangle $, which is an arbitrary unknown state. Then,
there is no universally unitary operator $W$ such that $W|\psi \rangle
=V|\varphi \rangle $, where $V$ is any universally unitary operator.

Proof. Assume that there exists a universally unitary operator $W$ such that
$W|\psi \rangle =V|\varphi \rangle $. Then, $(V^{-1}W)|\psi \rangle
=|\varphi \rangle $. It contradicts the fact that there exists no
universally unitary operator $U$ such that $U|\psi \rangle =|\varphi \rangle
$.

From Lemma 2 and Theorem 2, we have the following corollary.

\textit{Corollary 2. }There is no universally unitary operator$\ U$ such
that $U|\psi \rangle =V|\psi ^{\prime }\rangle $, where $|\psi \rangle
=\alpha |0\rangle +\beta |1\rangle $, which is an arbitrary unknown state, $%
V $ is any universally unitary operator, and $|\psi ^{\prime }\rangle $ is
one of $|\Phi _{i}\rangle $, $i=1,...,6$ in Table 2.

From Corollary 2, we obtain the general no-conjugating theorems from $V|\Phi
_{i}\rangle $, $i=1,...,6$, where $V$ is any universally unitary operator.
For example, we can obtain the general no-conjugating theorems from $\alpha
^{\ast }e^{i\theta }|0\rangle +\beta ^{\ast }e^{i\omega }|1\rangle $ and $%
\beta ^{\ast }e^{i\theta }|0\rangle +\alpha ^{\ast }e^{i\omega }|1\rangle $.

For example, by Corollary 2 there is no universally unitary operator $U$
such that $U|\psi \rangle =$ $|\Theta \rangle $, where $|\psi \rangle
=\alpha |{}0\rangle +\beta |1\rangle $, which is an arbitrary unknown state,
and $|\Theta \rangle =\sigma _{z}|\psi ^{\ast }\rangle =\alpha ^{\ast
}|0\rangle -\beta ^{\ast }|1\rangle $. So, we can call the no-go theorem the
no-conjugating the phase-flip theorem.

Specially, for the state $|\psi \rangle =\alpha |0\rangle +\beta |1\rangle $%
, let $|\psi ^{\bot }\rangle =\beta ^{\ast }|0\rangle -\alpha ^{\ast
}|1\rangle $ or $-\beta ^{\ast }|0\rangle +\alpha ^{\ast }|1\rangle $. One
can see that $|\psi \rangle $ and $|\psi ^{\bot }\rangle $ are orthogonal. $%
|\psi ^{\bot }\rangle $ is called the flip state of $|\psi \rangle $. Note
that $\beta ^{\ast }|0\rangle -\alpha ^{\ast }|1\rangle =\sigma _{z}|\Phi
_{6}\rangle $ and $-\beta ^{\ast }|0\rangle +\alpha ^{\ast }|1\rangle
=\sigma _{x}\sigma _{z}|\Phi _{3}\rangle $. By Corollary 2, we obtain
no-flip theorem, which states that no universally unitary operator can flip
an arbitrary unknown quantum state. Thus, it gives a different proof of
no-flip theorem. No-flip theorem was first proposed in \cite{bu, bu00}.

\section{Optimal fidelity of imperfect conjugating operation}

\subsection{The Bloch vector of $\protect\rho =|\protect\psi \rangle \langle
\protect\psi |$}

For $|\psi \rangle =\alpha |{}0\rangle +\beta |1\rangle $, where $|\alpha
|^{2}+|\beta |^{2}=1$,
\begin{equation}
\rho =|\psi \rangle \langle \psi |=\frac{1}{2}(I+\gamma .\sigma ),
\label{bloch}
\end{equation}%
which is simply the Bloch-sphere representation of a qubit density matrix,
where $I$ is the identity, $\sigma =\left(
\begin{array}{ccc}
\sigma _{x} & \sigma _{y} & \sigma _{z}%
\end{array}%
\right) ^{T}$, where $\sigma _{x},\sigma _{y},\sigma _{z}$ are Pauli
operators, and $\gamma =\left(
\begin{array}{ccc}
\gamma _{x} & \gamma _{y} & \gamma _{z}%
\end{array}%
\right) ^{T}$, which is called the Bloch vector.

For $|\psi \rangle =\alpha |0\rangle +\beta |1\rangle $, a calculation
yields
\begin{eqnarray}
\gamma _{x} &=&2\ast re(\alpha \beta ^{\ast })  \label{bloch-1} \\
\gamma _{y} &=&-2\ast im(\alpha \beta ^{\ast })  \label{bloch-2} \\
\gamma _{z} &=&|\alpha |^{2}-|\beta |^{2},  \label{bloch-3}
\end{eqnarray}%
where re(x) stands for the real part of x while im(x) for the imaginary part
of x.

Clearly, $\gamma _{x},\gamma _{y},\gamma _{z}$ in Eqs. (\ref{bloch-1}, \ref%
{bloch-2}, \ref{bloch-3})\ \ are real and $\gamma _{x}^{2}+\gamma
_{y}^{2}+\gamma _{z}^{2}=1$. One can check that Eq. (\ref{bloch}) holds via
Eqs. (\ref{bloch-1}, \ref{bloch-2}, \ref{bloch-3}).

\subsection{Optimal fidelity of imperfect conjugating operation}

For the conjugate state\ $|\psi ^{\ast }\rangle $\ of $|\psi \rangle $, a
calculation yields
\begin{equation}
\rho ^{\ast }=|\psi ^{\ast }\rangle \langle \psi ^{\ast }|=\frac{1}{2}%
(I+\gamma _{x}\sigma _{x}-\gamma _{y}\sigma _{y}+\gamma _{z}\sigma _{z})
\label{conj-2}
\end{equation}%
Therefore, for $\rho ^{\ast }=|\psi ^{\ast }\rangle \langle \psi ^{\ast }|$,
the Bloch vector is $\gamma =(\gamma _{x},-\gamma _{y},\gamma _{z})^{T}$.
The desired universal conjugating operation would transform in terms of the
Bloch vector

\begin{equation}
(\gamma _{x},\gamma _{y},\gamma _{z})\longrightarrow (\gamma _{x},-\gamma
_{y},\gamma _{z}).  \label{conj-3}
\end{equation}%
So universal conjugating operation corresponds to a reflection of the Bloch
sphere through the $xz$-plane. The exact universal reflection of the Bloch
sphere through the $xz$-plane.in Eq. (\ref{conj-3}) cannot be realized by a
universal unitary operation.

The optimal physical approximation can instead be chosen as a channel whose
output Bloch vector is

\begin{equation}
\gamma _{out}=\frac{1}{3}(\gamma _{x},-\gamma _{y},\gamma _{z}).
\end{equation}%
Note that the factor 1/3 is the largest value keeping the overall operation
completely positive and trace preserving. Thus,
\begin{equation}
\rho _{out}=\frac{1}{2}[I+\frac{1}{3}(\gamma _{x}\sigma _{x}-\gamma
_{y}\sigma _{y}+\gamma _{z}\sigma _{z})].
\end{equation}

The fidelity is defined as
\begin{equation}
F=Tr(\rho ^{\ast }\rho _{out}).
\end{equation}%
A calculation yields $F=2/3$, which does not depend on $|\psi \rangle $, so
the fidelity is universal.

\section{Optimal fidelity of imperfectly conjugating the bit-flip operation}

For $|\Phi _{6}\rangle =\beta ^{\ast }|{}0\rangle +\alpha ^{\ast }|1\rangle $%
, a calculation yields that
\begin{equation}
\rho ^{(6)}=|\Phi _{6}\rangle \langle \Phi _{6}|=\frac{1}{2}(I+\gamma
_{x}\sigma _{x}+\gamma _{y}\sigma _{y}-\gamma _{z}\sigma _{z})
\end{equation}%
Therefore, for $\rho ^{(6)}=|\Phi _{6}\rangle \langle \Phi _{6}|$, the Bloch
vector is $\gamma =(\gamma _{x},\gamma _{y},-\gamma _{z})^{T}$. The desired
universal conjugating the bit-flip operation would transform in terms of the
Bloch vector
\begin{equation}
(\gamma _{x},\gamma _{y},\gamma _{z})\longrightarrow (\gamma _{x},\gamma
_{y},-\gamma _{z}).  \label{xy-p}
\end{equation}

So universal conjugating the bit-flip operation\ corresponds to a reflection
of the Bloch sphere through the $xy$-plane. The exact universal reflection
of the Bloch sphere through the $xy$-plane in Eq. (\ref{xy-p}) cannot be
realized by a universal unitary operation.

The optimal physical approximation can instead be chosen as a channel whose
output Bloch vector is

\begin{equation}
\gamma _{out}=\frac{1}{3}(\gamma _{x},\gamma _{y},-\gamma _{z}).
\end{equation}%
Thus,%
\begin{equation}
\rho _{out}=\frac{1}{2}[I+\frac{1}{3}(\gamma _{x}\sigma _{x}+\gamma
_{y}\sigma _{y}-\gamma _{z}\sigma _{z})].
\end{equation}

A calculation yields the fidelity:
\begin{equation}
F=Tr(\rho ^{(6)}\rho _{out})=2/3,
\end{equation}%
which does not depend on $|\psi \rangle $, so the fidelity is universal.

\section{Optimal fidelity of imperfectly conjugating the phase-flip operation%
}

For\ the state $|\Theta \rangle =\alpha ^{\ast }|{}0\rangle -\beta ^{\ast
}|1\rangle $, a calculation yields
\begin{equation}
\rho _{\Theta }=|\Theta \rangle \langle \Theta |=\frac{1}{2}(I-\gamma
_{x}\sigma _{x}+\gamma _{y}\sigma _{y}+\gamma _{z}\sigma _{z})
\end{equation}%
Therefore, for $\rho _{\Theta }=|\Theta \rangle \langle \Theta |$, the Bloch
vector is $\gamma =(-\gamma _{x},\gamma _{y},\gamma _{z})^{T}$. The desired
universal conjugating the phase-flip operation would transform in terms of
the Bloch vector

\begin{equation}
(\gamma _{x},\gamma _{y},\gamma _{z})\longrightarrow (-\gamma _{x},\gamma
_{y},\gamma _{z}).  \label{yz-p}
\end{equation}%
So, conjugating the phase-flip operation corresponds to a reflection of the
Bloch sphere through the $yz$-plane. The exact universal reflection of the
Bloch sphere through the $yz$-plane in Eq. (\ref{yz-p}) cannot be realized
by a universal unitary operation.

The optimal physical approximation can instead be chosen as a channel whose
output Bloch vector is

\begin{equation}
\gamma _{out}=\frac{1}{3}(-\gamma _{x},\gamma _{y},\gamma _{z}).
\end{equation}%
Note that the factor 1/3 is the largest value keeping the overall operation
completely positive and trace preserving. Thus,
\begin{equation}
\rho _{out}=\frac{1}{2}[I+\frac{1}{3}(-\gamma _{x}\sigma _{x}+\gamma
_{y}\sigma _{y}+\gamma _{z}\sigma _{z})].
\end{equation}

The fidelity is defined as
\begin{equation}
F=Tr(\rho _{\Theta }\rho _{out}).
\end{equation}%
A calculation yields $F=2/3$, which does not depend on $|\psi \rangle $, so
the fidelity is universal.

Remark 1. If only one of two amplitudes $\alpha $ and $\beta $\ of the state
$|\psi \rangle =\alpha |{}0\rangle +\beta |1\rangle $ is conjugated, then we
say that we partially conjugate the state $|\psi \rangle =\alpha |{}0\rangle
+\beta |1\rangle $. We can show that the partially conjugating operations
including conjugating the first amplitude,\ conjugating the second
amplitude, conjugating the first amplitude of the bit-flip, and conjugating
the second amplitude of the bit-flip are not anti-unitary.

\section{Differences between no-flip theorem and no-conjugating theorem}

(1). $|\psi \rangle $ and its flip state $|\psi ^{\bot }\rangle $ are
orthogonal while $|\psi \rangle $ and its conjugate $|\psi ^{\ast }\rangle $
are not orthogonal.

(2). Via Bloch-sphere, the flip operation is the inversion through the
origin and the conjugating operation is the reflection across the $xz$-plane
\cite{bu, bu00}. We also indicate that conjugating the bit-flip (the
phase-flip) operation is the reflection across the $xy$-plane ($yz$-plane) .

(3). It was indicated that the flipping operator and the conjugating
operator\ are anti-unitary \cite{bu, bu00}. We indicate that conjugating the
bit-flip and phase-flip are also anti-unitary while the partially
conjugating operations are not anti-unitary.

\section{Implications for Physics}

(1). No conjugating gate for quantum circuit.

It establishes a fundamental limitation on what quantum gates can do. A gate
cannot universally conjugate an arbitrary unknown qubit.

(2). Quantum cryptography

No-conjugating theorem helps explain why an eavesdropper cannot simply take
an unknown quantum state and convert it into its conjugate to obtain
information or distinguish quantum states perfectly.

\section{Conclusions}

In this paper, we have the following conclusions.

(1). It is impossible to conjugate or reflect an unknown state via a
universal unitary operation.

(2). It is impossible to conjugate or reflect the bit-flip of an unknown
state via a universally unitary operation.

(3). It is impossible to conjugate or reflect the phase-flip of an unknown
state via a universally unitary operation.

(4). It is impossible to partially conjugate an unknown state, the bit flip
of an unknown state, or the phase flip of an unknown state via a universally
unitary operation.

For example, there is no universally unitary operator $U$ such that $U|\psi
\rangle =|\psi ^{\prime }\rangle $, where $|\psi \rangle =\alpha |0\rangle
+\beta |1\rangle $, which is an arbitrary unknown state, and $|\psi ^{\prime
}\rangle =\alpha |0\rangle \pm \beta ^{\ast }|1\rangle $, $\alpha ^{\ast
}|0\rangle \pm \beta |1\rangle $, $\beta |0\rangle \pm \alpha ^{\ast
}|1\rangle $, or $\beta ^{\ast }|0\rangle \pm \alpha |1\rangle $.

We compute the fidelities of the optimal physical approximations to
conjugating, conjugating the bit-flip, and conjugating the phase-flip
operations.

Acknowledgement. Thank Prof. I. Chakrabarty for his useful comments and
suggestions.

Statements and declarations: No financial interests, no competing interests,
no financial supports.

Data Availability Statement: No Data associated in the manuscript.


\begin{thebibliography}{99}
\bibitem{Wootters} W. K. Wootters and W. H. Zurek., A Single quantum cannot
be cloned. Nature 299, 802-803 (1982).

\bibitem{Yao} Yuan Yao,\ Phase spaces that cannot be cloned in classical
mechanics, J. Math. Phys. 64, 102901 (2023).

\bibitem{Yama} Koji Yamaguchi and Achim Kempf, Encrypted Qubits can be
Cloned, Phys. Rev. Lett. 136, 010801 (2026)

\bibitem{Pati} A.K. Pati and S.L. Braunstein, Impossibility of deleting an
unknown quantum state.\ Nature 404, 164 (2000). arxiv, e-print 9911090v2.

\bibitem{Qiu} D. Qiu, Some analogies between quantum cloning and quantum
deleting, Phys. Rev. A 65, 052303 (2002)

\bibitem{sk} S. Sazim et al., Complementarity of Quantum Correlations in
Cloning and Deleting of Quantum State, Phys. Rev. A 91, 062311 (2015).

\bibitem{IC} I. Chakrabarty et al., Deletion, Bell's Inequality,
Teleportation, Quant. Inf. Process, 10, 27 (2011).

\bibitem{Chen} Y. Chen et al., Quantum deleting and cloning in a
pseudo-unitary system, Front. Phys. 16, 53601 (2021)

\bibitem{Samuel} Braunstein, Samuel L. and Pati, Arun K., Quantum
Information Cannot Be Completely Hidden in Correlations: Implications for
the Black-Hole Information Paradox. Phys. Rev. Lett. 98 (8), 080502 (2007).

\bibitem{Barnum} Barnum, Howard; Caves, Carlton M.; Fuchs, Christopher A.;
Jozsa, Richard; Schumacher, Benjamin. \ Noncommuting Mixed States Cannot Be
Broadcast. Phys. Rev. Lett..76 (15), 2818--2821 (1996).
arXiv:quant-ph/9511010.

\bibitem{Matt} Matthew Girling, Cristina C\^{\i}rstoiu, and David Jennings,
A simple formulation of no-cloning and no-hiding that admits efficient and
robust verification,\ Phys. Rev. Research 6, 023090 (2024)

\bibitem{bu} Bu\v{z}ek, M. Hillery, and R.F. Werner, Optimal Manipulations
with Qubits: Universal NOT Gate, Physical Review A 60, R2626--R2629 (1999),
e-print, arxiv/quant-ph/9901053

\bibitem{bu00} V. Bu\v{z}ek, M. Hillery, and R. F. Werner, Universal-NOT
gate, Journal of Modern Optics 47, 211--232 (2000).

\bibitem{Pati05} A. K. Pati and B. C. Sanders, No partial erasure of quantum
information, Physics Letters A 359, 31-36 (2006)

\bibitem{IC-2} I. Chakrabarty, Impossibility of partial swapping of quantum
information, International Journal of Quantum Information Vol. 05, No. 04,
pp. 605-609 (2007)
\end{thebibliography}
\end{document}